\documentclass[twocolumn,preprintnumbers,amsmath,amssymb10pt]{revtex4}

\usepackage{graphicx}
\usepackage{dcolumn}
\usepackage{bm}
\usepackage{color}
\usepackage{lscape}
\usepackage{wrapfig}

\begin{document}

\title{Superstability of the yeast cell cycle dynamics:\\ 
Ensuring causality in the presence of biochemical stochasticity}    
\author{Stefan Braunewell}
\author{Stefan Bornholdt}
\email{bornholdt@itp.uni-bremen.de}
\affiliation{ {Institute for Theoretical Physics, University of Bremen, 
D-28359 Bremen, Germany}
}  
\bibliographystyle{apalike}
\begin{abstract}
Gene regulatory dynamics is governed by molecular processes and 
therefore exhibits an inherent stochasticity. However, for the survival 
of an organism it is a strict necessity that this intrinsic noise does not 
prevent robust functioning of the system. It is still an open question 
how dynamical stability is achieved in biological systems despite 
the omnipresent fluctuations. 
In this paper we investigate the cell-cycle of the budding yeast 
Saccharomyces cerevisiae as an example of a well-studied organism. 

We study a genetic network model of eleven genes that 
coordinate the cell-cycle dynamics
using a modeling framework which generalizes the concept of 
discrete threshold dynamics. By allowing for fluctuations in the 
transcription/translation times, we introduce noise
in the model, accounting for the effects of biochemical stochasticity. 
We study the dynamical attractor of the cell cycle and find a remarkable 
robustness against fluctuations of this kind. We identify mechanisms that 
ensure reliability in spite of fluctuations:  `Catcher' states and persistence of 
activity levels contribute significantly to the stability of the yeast cell cycle 
despite the inherent stochasticity.
\\
\begin{center}
\end{center}
\end{abstract}

\maketitle
\section{Introduction}
Uncovering the complex mechanisms involved in gene regulation remains to be a 
major challenge. 
While the biochemical processes involved in the expression of a single gene are 
increasingly well understood, the interplay of whole networks of genes poses 
additional questions and it is unclear what level of system-specific detail has 
to be taken into account to describe a gene regulatory network \cite{SBB2000,Ty2001}. 
One promising approach to modeling system wide dynamical states of a network 
is to go to an abstract level of description \cite{Bornholdt:2005tc} which may even include 
discrete deterministic models such as Boolean or threshold networks 
\cite{Ka1969,Thomas1973}. We here extend a recent model of the yeast cell cycle 
dynamics that is successfully based on this approach \cite{Li2004}. 

The cell cycle of the budding yeast Saccharomyces cerevisiae is a widely studied 
example of a robust dynamical process \cite{Sp1998,Lee:2002}. 
In \cite{Li2004}, the yeast cell cycle was modeled in the framework of a discrete 
threshold network. From the data in \cite{Sp1998}, eleven genes that play 
a key role in the cell cycle process were identified along with their known 
(direct or indirect) interactions. The activity of a certain gene is modeled as a 
two-state system, with values 1 (active) or 0 (inactive). Using a threshold model 
of interactions, the biological sequence of activity states in the process is exactly 
reproduced. The authors also find considerable dynamical robustness properties 
that can be traced to the properties of the basin of attraction of the biological fixed point.

Remarkably, these results were obtained using a discrete time model, where 
each discrete time step $t$ is defined by the intervals between activity changes.
In the model, the activity state of every gene is determined solely by the state of its 
transcription factors at the previous time step. It is remarkable, that in this case at least, 
the biochemical stochasticity of gene regulation \cite{McAdams:1997,Rao:2002fk} 
can be neglected in the model.
 
In particular, as was shown in \cite{KB2004-2}, attractors under synchronous 
dynamics can be unstable if stochasticity is imposed on the transmission times. 
In this work, we investigate whether the cell-cycle process is stable under such perturbations.

Investigations of dynamical robustness have been discussed in a variety of 
different biological systems, such as segmentation in the fruit fly 
\cite{vDMMO2000,AO2003,Chaves:2005lr}, or  two-gene circadian 
oscillators \cite{Wagner:2005}. 
Different conceptions of the word `robustness' have been used \cite{Kitano2004}. 
Robustness against mutations means that a specific process can be performed
reliably by a system even if some changes to the structure of the system are conducted. 
The yeast cell-cycle is remarkably robust in this sense \cite{Li2004}.
Other approaches to assessing robustness in biological networks include local 
stability and bifurcation analyses \cite{Chen:2002tn}, stability under node state 
perturbation \cite{Aldana:2003,KPST2004} and probabilistic Boolean networks 
\cite{Shmulevich:2002}. In this work we will concentrate on the robustness under 
stochastically varying processing times (for protein concentration buildup and decay) 
as was considered in \cite{KB2004}.

Other models of the yeast cell-cycle include molecular models of major CDK 
activities in Start and Finish states \cite{Chen01012000} and of S-phase entrance 
in \cite{ARPV2005}. In \cite{Chen2005} stochastic differential equations have been 
used to fit time-courses of protein concentration levels in the yeast cell-cycle network.

\section{Model description}
\renewcommand{\tabcolsep}{0.14cm}

\begin{table*}
\footnotesize
\begin{tabular}{ccccccccccccc} 
\hline
Time & Cln3 & MBF & SBF & Cln1,2 & Cdh1 & Swi5 & Cdc20/ & Clb5,6 & Sic1 & Clb1,2 & Mcm1/ & Phase \\
& & & & & & &Cdc14 & & & &SFF &\\\hline
\bf \textcolor{blue} 1 &\bf \textcolor{blue} 1 & 0 & 0 & 0 & 1 & 0 & 0 & 0 & 1 & 0 & 0 & Start \\
\bf \textcolor{red}2 & 0 & \bf \textcolor{red}1 &\bf \textcolor{red} 1 & 0 & 1 & 0 & 0 & 0 & 1 & 0 & 0 & $G_1$\\
\bf \textcolor{blue}3 & 0 & 1 & 1 & \bf \textcolor{blue}1 & 1 & 0 & 0 & 0 & 1 & 0 & 0 & $G_1$\\
\bf \textcolor{red}4 & 0 & 1 & 1 & 1 & \bf \textcolor{red}0 & 0 & 0 & 0 & \bf \textcolor{red}0 & 0 & 0 & $G_1$\\
\bf \textcolor{blue}5 & 0 & 1 & 1 & 1 & 0 & 0 & 0 & \bf \textcolor{blue}1 & 0 & 0 & 0 & S \\
\bf \textcolor{red}     6 & 0 & 1 & 1 & 1 & 0 & 0 & 0 & 1 & 0 & \bf \textcolor{red}1 & \bf \textcolor{red}1 & $G_2$\\
\bf \textcolor{red}     7 & 0 &\bf \textcolor{red}      0 & \bf \textcolor{red}     0 & 1 & 0 &  0 & \bf \textcolor{red}    1 & 1 & 0 & 1 & 1 & M\\
\bf \textcolor{red}     8 & 0 & 0 & 0 & \bf \textcolor{red}     0 & 0 & \bf \textcolor{red}    1 & 1 & \bf \textcolor{red}    0 & 0 & 1 & 1 & M\\
\bf \textcolor{blue}     9 & 0 & 0 & 0 & 0 & 0 & 1 & 1 & 0 &\bf \textcolor{blue}      1 & 1 & 1 & M\\
\bf \textcolor{blue}     {10} & 0 & 0 & 0 & 0 & 0 & 1 & 1 & 0 & 1 &\bf \textcolor{blue}      0 & 1 & M\\
\bf \textcolor{red}    {11}  & 0 & 0 & 0 & 0 & \bf \textcolor{red}     1 & 1 & 1 & 0 & 1 & 0 &\bf \textcolor{red}      0 & M\\
\bf \textcolor{blue}     {12} & 0 & 0 & 0 & 0 & 1 & 1 & \bf \textcolor{blue}     0 & 0 & 1 & 0 & 0 & $G_1$\\
\bf \textcolor{blue}     {13} & 0 & 0 & 0 & 0 & 1 & \bf \textcolor{blue}     0 & 0 & 0 & 1 & 0 & 0 & $G_1$\\\hline
\end{tabular}
\caption{The synchronous sequence of states as recorded in \cite{Li2004}. 
We color time steps in which only one switch occurs blue, and those with more than one switch red.}
\label{tsequence}
\end{table*}

Following \cite{Li2004}, a network of eleven nodes is used to describe the cell cycle process. 
They are given in table \ref{tsequence}, along with the synchronous sequence of activity 
states recorded in that work. Using a technique introduced in \cite{Glass:1975} we extend 
that model to include fluctuating transmission delays and to allow for 
real numbers for protein concentrations levels ($0\leq c_i(t) \leq 1$ for protein $i$).
We keep the characteristics of the description of \cite{Li2004}, that is the effect of protein 
$j$ on the transcription of protein $i$ is determined by a discrete activity state 
(`active' or `inactive') of protein $j$. In our continuous description, we set the 
activity state $S_i$ of a protein to 1 if the concentration is above a certain 
threshold ($c_i(t)>0.5$), otherwise it is 0.

The transmission function that determines the transcription or degradation of protein $i$ is given by
\begin{equation}\label{updaterule}
f_i(t,t_d)=\left \{ 
\begin{array}{lll}
1,\quad &\sum_j a_{ij}S_j(t-t_d)>0,\\
0,\quad &\sum_j a_{ij}S_j(t-t_d)<0.
\end{array} \right.
\end{equation}
where $t_d$ is the transmission delay time that comprises the time taken by processes such as translation or diffusion that cause the concentration buildup of one protein to not immediately affect  other proteins. The numbers $a_{ij}$ determine the effect that protein $j$ has on protein $i$. An activating interaction is described by $a_{ij}=1$, inhibition by $a_{ij}=-1$. If the presence of protein $j$ does not affect expression of protein $i$, $a_{ij}=0$.

If $\sum_j a_{ij}S_j(t-t_d)=0$, the value of $f_i$ depends on whether the node is modeled as a self-degrader. Self-degraders are those nodes that are down-regulated by external processes (Cln3, Cln1,2, Swi5, Cdc20/Cdc14,  Mcm1/SFF). Self-degrader nodes will take a value $f_i(t,t_d)=0$ whereas the transmission function of non-self-degraders is left unchanged, i.e.~the last time $\tilde t$ when $f_i(\tilde t,t_d)\neq 0$ determines the state at time $t$.

We now describe the time evolution of the system of genes by the following set of delay differential equations
\begin{equation}
\frac{dc_i(t)}{dt}
=f_i(t,t_d)-
\frac{c_i(t)}{\tau}.
\end{equation}

For the simple transmission function given above, this equation can be easily solved piecewise (for every period of constant transmission function), leading to charging behavior of the concentration levels
\begin{equation}\label{solution}
c_i(t>t_0)=\left\{
\begin{array}{ll}
	1-(1-c(t_0))\exp(-(t-t_0)/\tau) & f_i \geq 0, \\
	c(t_0) \exp(-(t-t_0)/\tau) & f_i < 0.
\end{array}\right.
\end{equation}

\begin{figure}
\begin{center}
\vspace{-0.5cm}
\includegraphics[width=8.5cm]{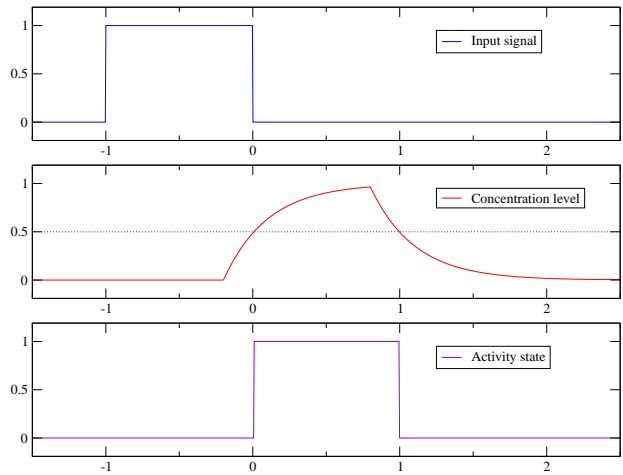}
\caption{Concentration buildup and decay of a protein given a specific input signal and the corresponding activity state ($t_d=0.8$, $\tau=0.3$).}
\label{fcharging}
\vspace{-0.5cm}
\end{center}
\end{figure}

This has the effect of a low-pass filter, 
i.e., 
a signal has to sustain for a while to affect the discrete activity state. A signal spike, on the other hand, will be filtered out. Concentration buildup in our model is depicted in figure \ref{fcharging}. Here, the transcription factor of a protein is assumed to be present in the time span $t=-1$ to $0$ (upper panel). The production of the protein starts after the delay time (here $t_d=0.8$) and the concentration crosses the critical level of $0.5$ at $t=1$ (central panel), switching the activity state to ``on'' (lower panel). 
In the case of very fast build-up and decay ($\tau \to 0$ in eq. (\ref{solution})) and with the delay time set to one ($t_d=1$), we exactly recover the synchronous dynamics of \cite{Li2004}. Thus, our described model is a simple generalization of the synchronous case
to allow for a continuous time description.

We now ask the following question: 
Is the original sequence stable 
under stochastic timing noise (stochastically varying signal delay times)
or can the noise cause the system to assume different states? As the sequence from \cite{Li2004} (reproduced in table \ref{tsequence}) runs into the stationary $G_1$ fixed point and an external signal is needed to trigger the starting state again, we create a repeating cycle of states (limit cycle) by explicitly adding the rule that Cln3 production is triggered as soon as the final state in the synchronous sequence is reached. We will investigate whether this limit cycle is inherently stable or whether it needs the perfect synchronization of the artificial synchronous update.

In this context it is important to note that the stability of the complete cell-cycle system also depends on the behavior of all other proteins involved. However, the stability of the core genes is most important, as they regulate the other proteins. Only if the regulators perform reliably, the system as a whole can be robust.

To compare the time series of our simulations with the discrete time steps of the synchronous case, we record a time step whenever the system keeps all its activity states constant for a time span of at least $t_d/2$. With every switch of activity states (say, at time $t_0$) we check whether the transcription of any other protein P is affected. If so, the concentration level of protein P will begin to rise at time $t_0 + t_d + \chi$ where $\chi$ denotes a uniformly distributed random number between $0$ and $\chi_\mathrm{max}$ that perturbs the delay times.

Our simulation time is not directly related to the actual time intervals of the biological processes involved. However, we are not so much interested 
in the specifics of the time course but rather in the properties of stability and for this assessment it is not important how long the actual phases take. Our model captures two principles of real world gene regulatory networks: Interactions occur with a characteristic time delay (denoted by $t_d$); and we use continuous concentration levels and implement low pass filter behavior due to protein concentration buildup with a characteristic time $\tau$ \cite{Hi2002}. 

\section{Results}

First, we check if the system reproduces the synchronous sequence under small perturbations of the delay time. Thus we stay in the regime where $\chi_\mathrm{max}$ is significantly smaller than the characteristic protein decay or buildup time $\tau$. In the main simulation runs we set $t_d=1$, $\tau=0.3$ and $\chi_\mathrm{max}=0.1$, but any numbers that fulfill $\chi_\mathrm{max}\ll t_d$ give the same results.

We find that the synchronous sequence of states is reliably reproduced by this stochastic dynamics. Even long simulation runs of $t<10^7$ cannot push the system out of the original attractor. This means that the biological sequence is absolutely stable against small perturbations.

To understand this, we look at the synchronous sequence of states in table \ref{tsequence}. In steps $2\to3$, $4\to5$, $8\to9$, $9\to10$, $11\to12$, $12\to13$ (marked blue in the table) only a single protein changes its activity state. If all steps were of this kind, fluctuations of the event times would not be able to destroy the attractor at all. States marked in red denote events where multiples switches happen at the same time.

To illustrate this point, let's assume two nodes switch their states at times $t_1$ and $t_2$ (we call this a `phase lag'). The system thus assumes an intermediate state in the time span between $t_1$ and $t_2$. Approximately at time $t_1 + t_d$ the next switches occur and due to the intermediate state it is possible that proteins switch their states which would normally be constant in this step. Because of the charging behavior of the concentration levels, these `spikes' will be filtered out. The only way to destroy the attractor is thus when the phase lags add up in a series of steps. This cannot happen in the yeast cycle, however, due to the states marked in blue color in the table. When only one protein changes its state in a time step, all divergence of signal times will be reset and the synchrony is restored. We therefore call these steps `catcher states' as they remove phase-lags from the system.

Now that we know that small perturbations cannot drive the system out of the synchronous attractor, we want to investigate stability under stronger noise. To address this question, we have to loosen our definition of stability. Up to now, we have requested the system to follow the exact sequence of states of the synchronous dynamics. It is clear that this strict stability cannot be obtained if we increase the noise to be more than half of the transmission delay itself, because two nodes switching at the same synchronous time step can receive switching times that differ by more than $t_d/2$. The intermediate step taken when only one node has switched obviously violates the stability criterion. 

To assess the stability of the system under strong noise, we employ a different stability criterion. We let the system run with the sole constraint that the stationary $G_1$ state will be assumed regularly for a time span of at least $t_d$. Any fluctuations occurring inbetween two $G_1$ incidences will be tolerated, as long as the system finds its way to the $G_1$ state of the cell cycle in which growth occurs until the cell size signal is triggered. Although this might seem too loose a criterion for robust biological functioning, one has to remember that the cell-cycle process is also backed up by a system of checkpoints that can catch faulty system states. We investigate here the inherent stability of the system disregarding these checkpoints but at the same time allowing more variability in the sequence.

\begin{figure}
\begin{center}
\includegraphics[width=8.5cm]{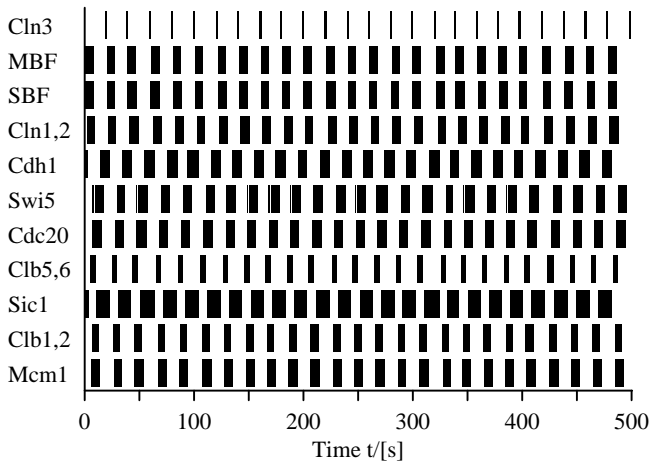}
\caption{
Time course of a run with noise of the order of the delay time $\chi_\mathrm{max}=0.9~t_d$. Black boxes denote active states, white means inactive. On a micro-time level the effect of fluctuations is visible, but on a larger time scale the dynamics is very stable.}
\label{stablenoise}
\end{center}
\end{figure}

Remarkably, with noise of the order of the delay time and largely independent of the filter used, the system reliably stays in the biological attractor. An example run with $t_d=1$, $\tau=0.3$ and $\chi=0.9$ ran for a time of $10^7$ following the biological attractor sequence (in the wider sense mentioned above). A typical time span of this run is shown in figure \ref{stablenoise}. This is a surprising result, because in general one expects a system to be able to leave its attractor sequence under such strong noise if a series of multi-switch events (steps 5 to 8) is involved anywhere during the sequence.

\begin{figure}
\begin{center}
\includegraphics[width=8.5cm]{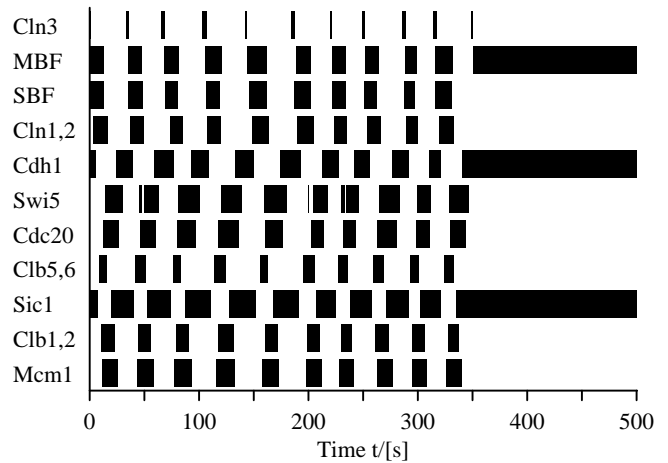}
\caption{Time course of an example run with strong noise $\chi_\mathrm{max}=3~t_d$. After some repetitions of the biological state sequence the attractor cycle is lost and a fixed point is assumed.}
\label{strongnoise}
\end{center}
\end{figure}

Our proposed criterion is not trivially fulfilled: by changes in the sequence of switching events or by delaying one of several events that occur at the same synchronous time step, a new sequence could be triggered. This could force the system to jump into one of the other six fixed points identified in \cite{Li2004} without the possibility to return to the biological sequence. In figure \ref{strongnoise} we show an example of a simulation run with extremely strong noise $\chi_\mathrm{max}=3~t_d$ that shows that the system can jump out of the attractor. However, it is also apparent that even under such strong fluctuations the system runs quite regularly until it finally loses its attractor sequence. 

\begin{figure}
\begin{center}
\includegraphics[angle=-90,width=8.5cm]{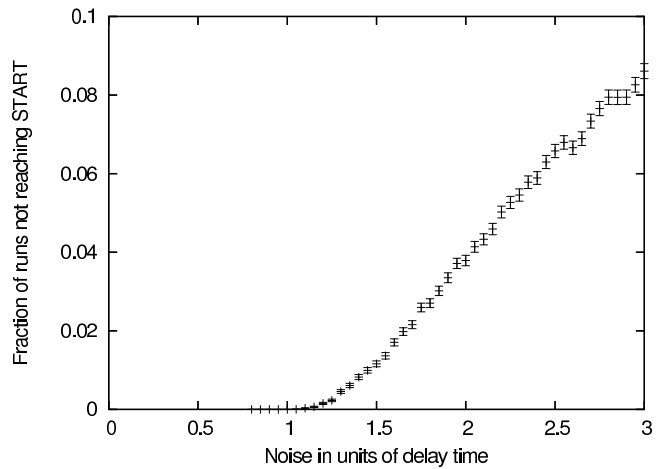}
\caption{The ratio of runs escaping the biological limit cycle plotted against the maximal noise level $\chi_\mathrm{max}$. Even for strong noise, the fraction of erroneous runs is very small.}
\label{errorsvsnoise}
\end{center}
\end{figure}

We now quantify the stability of the biological pathway under such strong noise. How likely is it for the system to lose its biological sequence and to run into a different fixed point? To address this question, we initialize the system at the Start state again and check whether it completes the cycle. Again, we use the lose criterion described above, which means we only request the system to reach the Start state again. In figure \ref{errorsvsnoise} we show the ratio of erroneous runs of the biological pathway plotted against the noise level $\chi_\mathrm{max}$. It can be clearly seen that for reasonable noise levels the ratio of sequence runs not ending in a biological fixed point is very small. In fact, even with unrealistically high noise levels of $\chi_\mathrm{max}=20$ or more (which amounts to arbitrary update times), only in a quarter of the runs the system jumps out of the biological state sequence.

The by far dominating cause for this (very small) instability is the first step (cf. table \ref{tsequence}) where both SBF and MBF are activated by Cln3. If the Cln3 concentration is degraded before activating the transcription of either SBF or MBF, the system loses the biological sequence. If we explicitly force Cln3 activity to sustain long enough to make sure that both SBF and MBF are produced, even this small instability vanishes and the system assumes practically complete stability for all reasonable noise levels ($0.1\%$ erroneous runs at $\chi_\mathrm{max}=3 t_d$). This superstability is due to the fact that all proteins keep their activity states for an extended time. Extremely strong noise is therefore needed to delay a single activity switch long enough to significantly perturb the system.

We have tested all results with a wide variety of parameters. With a fixed number for the delay time $t_d$, only the noise level $\chi_\mathrm{max}$ and the characteristic protein buildup time $\tau$ can be adjusted. Our results are completely robust against changes of $\tau$, even removing the filter completely or setting it an order of magnitude larger than the delay time does not affect the robustness properties described above. 

\section{Discussion}
As we have shown in the previous section, the yeast cell-cycle control network is astonishingly stable against fluctuations of the protein activation and degradation times.
The network and the resulting dynamics exhibit a number of features that cause this stability: As was already discussed in \cite{Li2004}, the basin of attraction is very large, making it unlikely that an intermediary state belongs to one of the other fixed point basins. A second remarkable property is that all node states are sustained for at least three (synchronous) steps, making the system less dependent on the specifics of the concentration buildup procedure. Third and most important for the observed superstability under noisy transmission times, is the presence of the catcher states which prevent the system from gradually running out of synchrony. 

Thus, we have seen that without even taking into account the biological checkpoint mechanisms that give additional stability and error-correction features, the system shows a strong inherent robustness against intrinsic fluctuations.
In this example of the yeast cell-cycle dynamics, potential mechanisms that provide robustness under biological noise can be observed. A system without an external clock (or any other external control) can still run reliably if it has intrinsic features that enforce robustness: catcher states, persistence of states and an attractor landscapes that minimizes the possibilities to escape the biological sequence.

To conclude, we have investigated the stability of the cell-cycle network by extending the model of Li et al.~ to allow asynchronous updating of the activity states of the genes. We find that the system exhibits  robust behavior under noisy transmission times. Even without taking into account the checkpoint mechanisms that give additional stability and fallback features, the system shows a strong inherent robustness that aids in maintaining reliable functioning.

\end{document}